\documentclass[sigconf]{acmart}

\usepackage{microtype}
\usepackage{booktabs} 
\usepackage[ruled,vlined,linesnumbered,resetcount]{algorithm2e}
\usepackage{arydshln}
\usepackage{subcaption}
\usepackage{hyperref}
\usepackage{flushend}
\hypersetup{colorlinks=true}

\setcopyright{rightsretained}

\begin{document}
\title[Vertex-Context Sampling for Weighted Network Embedding]
{Vertex-Context Sampling for Weighted Network Embedding}

%
%
%

\author{Chih-Ming Chen}
\authornote{Social Networks and Human-Centered Computing Program,
     Taiwan International Graduate Program,     
     Academia Sinica, Taipei, Taiwan}
\affiliation{%
  \institution{Department of Computer Science,\\National Chengchi University}
  \city{Taipei} 
  \country{Taiwan} 
}
\email{104761501@nccu.edu.tw}

\author{Ming-Feng Tsai}
\affiliation{%
  \institution{Department of Computer Science,\\National Chengchi University}
  \city{Taipei} 
  \country{Taiwan} 
}
\email{mftsai@nccu.edu.tw}

\author{Yian Chen}
\affiliation{%
  \institution{Research Center, KKBOX Inc.}
  \city{Taipei}
  \country{Taiwan} 
}
\email{annchen@kkbox.com}

\author{Yi-Hsuan Yang}
\affiliation{%
  \institution{Research Center for Information Technology Innovation}
  \city{Taipei}
  \country{Taiwan} 
}
\email{yang@citi.sinica.edu.tw}



\begin{abstract}

Network embedding methods have garnered increasing attention because of their effectiveness in various information retrieval tasks. The goal is to learn low-dimensional representations of vertexes in an information network and simultaneously capture and preserve the network structure. Critical to the performance of a network embedding method is how the edges/vertexes of the network is sampled for the learning process. Many existing methods adopt a uniform sampling method to reduce learning complexity, but when the network is non-uniform (\textit{i.e.} a weighted network) such uniform sampling incurs information loss. The goal of this paper is to present a generalized vertex sampling framework that works seamlessly with most existing network embedding methods to support weighted instead of uniform vertex/edge sampling. For efficiency, we propose a delicate sequential vertex-to-context graph data structure, such that sampling a training pair for learning takes only constant time. For scalability and memory efficiency, we design the graph data structure in a way that keeps space consumption low without requiring additional space. Moreover, the proposed framework can be used to implement extensions that feature high-order proximity modeling and weighted relation modeling. Experiments conducted on three datasets, including a commercial large-scale one, verify the effectiveness and efficiency of the proposed weighted network embedding methods on various tasks, including word similarity search, multi-label classification, and item recommendation.

\end{abstract}

%
%


\maketitle

\section{Introduction}


Nowadays, information networks are everywhere in the real world, such as publication networks, social networks, and the user-item preference networks in recommender systems. Network embedding is an important unsupervised method to learn the low-dimensional representations of vertexes in an information network, the purpose of which is to capture and preserve the network structure. Such a low-dimensional representation has been shown useful in various information retrieval and data mining applications such as node classification, link prediction, and recommendation~\cite{hne}.

Existing network embedding methods differ in a number of aspects regarding how they learn the representations of vertexes of a network ~\cite{wl,com,hpe,grarep,app}. DeepWalk~\cite{dw} is one of the pioneering studies that adapts the concept of the skip-gram model to learn the vertex representations of a uniform graph. This is done based on local information obtained from a truncated random walk over the vertexes.  LINE~\cite{line} proposes a method that claims to preserve both information of local and global network structures.
It presents a method to deal with weighted edges, but it considers only the directly connected relations.
Node2vec~\cite{n2v} further demonstrates that breadth-first/depth-first search walking style can better explore the interconnected relations among vertexes. Recent years witnessed increasing interest in using deep learning techniques for  representation learning~\cite{sdne,hne,dngr}, but these studies still share the same idea of learning the representations from a network.

To preserve the network structure by the low-dimensional representation, the key modeling step is to determine which vertexes are sampled to be considered in the learning process.  The idea is that similar vertexes share similar connections. In general, the learning process consists of two steps: 1) draw the predefined positive/negative relations; and 2) update the representations by optimizing the predefined objective function.  By iteratively processing the two steps, the learning process is expected to obtain the appropriate vertex representations.

It is usually not possible to consider all the positive and negative relations in the data for the first step of the learning process, for storing all the pair-wise relations demands too much space for a large information network. For scalability, some sort of sampling for the training data is needed.  In previous studies, the uniform random sampling is often utilized because of its simplicity~\cite{dw,wl,hpe,com}.  However, uniform sampling ignores the strength of relations (i.e. weights) among vertexes.  We consider the relation strength as one important feature because it affects the distribution of learned representations~\cite{dwmf}. Although some efforts have been made to achieve weighted sampling, such as those presented in LINE~\cite{line} and node2vec~\cite{n2v}, these methods suffer from at least one of following issues: 1) the sampling requires a high computation cost, 2) it takes a great amount of memory space and 3) the sampling is restricted to first-order proximity of a vertex (i.e. directly connected relations) and cannot exploit high-order proximities.





This paper attempts to overcome the above issues and seeks to further improve the modeling quality and flexibility of the existing network embedding methods, while keeping space and time complexity low.  Specifically, we propose a novel sequential vertex-context graph data structure that stores the connections status in an sequential way, and use the data structure to realize a new weighted vertex-context sampling method.
Our sampling is efficient because, with the new data structure, the entire network can be divided into multiple sub-networks based on certain characteristics.  We make use of this property to present weighted vertex sampling functions for vertex-context sampling.
In addition to implementing existing network embedding methods over a weighted information network, we show that the proposed framework can be used to implement extensions that feature high-order proximity modeling and weighted relation modeling.

In our experiments, two public benchmarks and one on-line evaluation with commercial data are used to verify the effectiveness and efficiency of the proposed method. Experimental results show that the proposed method outperforms several state-of-the-art methods on the benchmark datasets on a variety of tasks, including similarity search, multi-label classification and item recommendation.  

The main contributions of this paper are summarized as follows:
\begin{itemize}
  \item We develop a generalized weighted network embedding framework that can support the implementation of various network embedding methods and their extensions, for a large-scale weighted information network. 
  \item The proposed weighted vertex-context sampling method and the sequential vertex-context  graph data structure helps keep the time and space complexity low.
  \item The experiments are conducted on three large-scale datasets, including one language network and two preference networks.  The results validate the effectiveness and efficiency of the presented weighted versions of various network embedding methods.
  \item We release our source code for reproducibility and for promoting follow-up research along this line.
  ~\footnote{\href{https://github.com/cnclabs/proNet-core}{https://github.com/cnclabs/proNet-core}}
\end{itemize}

In what follows, we present in Section \ref{sec_problem_formulation} a problem formulation of weighted network embedding, in Section \ref{sec_related} a review of existing non-uniform sampling techniques, and in Section \ref{sec_proposed} the proposed data structure and vertex-context sampling method.
We demonstrate in Section \ref{sec:wne} how the proposed framework can be used to support the implementation of various network embedding methods and their variants, and in Section \ref{sec:exp} experiments that validate the effectiveness of our implementations.
Section \ref{sec_conclusion} concludes the paper.

\section{Problem Formulation}
\label{sec_problem_formulation}


A common underlying assumption of a network embedding method is that the vertexes sharing the similar set of proximities are embedded (projected) into vectors having similar localities in the latent space~\cite{line,dw,hpe,n2v,wl,grarep}. Given a vertex $v_{i}$, the probability of observing a vertex $v_{j}$ as its neighbor can be defined as modeling the following conditional probabilities:
\begin{equation}
    p(v_{j}|v_{i}) = \frac{sim(v_{i}, v_{j})}{\sum_{v_{k}}{sim(v_{i},v_{k})}},
    \label{eq:context}
\end{equation}
where $sim(\cdot,\cdot)$ is a custom function that measures the similarity between the two vertexes $v_{i}$ and $v_{j}$. The main difference among existing network embedding methods usually lies in the way this probability distribution is defined and computed.

\subsection{Sampling-based Learning Framework}

In this work, we propose to model the similarities by setting the ground truth of a vertex pair $(v_{i}, v_{j})$ as:
\begin{equation}
    p(v_{i}, v_{j}) =
    \begin{cases}
        1 & \text{if } v_{j} \in Proximities(v_{i}) \\
        0 & \text{otherwise}
    \end{cases}.
    \label{eq:gt}
\end{equation}
In this case, adopting different $Proximities$ sets will change the similarities among the vertexes in a network, leading to different probability distributions.  
Accordingly, the proposed framework can be used to implement various existing network embedding methods by sampling different vertex pairs for $Proximities$.



To learn vertex representations in a flexible way, the built estimation function $\hat{p}$ is modeled by two representation mapping functions, $\Phi: v \mapsto \mathbb{R}^{d}$ and $\Phi': v \mapsto \mathbb{R}^{d}$. Each of them converts a vertex $v$ into a $d$-dimensional vector, so that the mapping functions are adjustable towards certain assumptions of the network embedding methods:
\begin{equation}
\hat{p}(v_{j}|v_{i}) = \sigma( \Phi(v_{i}) \cdot \Phi'(v_{j}) ),
\label{eq:p}
\end{equation}
where $\sigma(\cdot)$ is a sigmoid function rescaling the values to [0,1]. For instance, a conventional network embedding solution is to treat each vertex as having two roles: 1) the ``vertex'' itself (\textit{i.e.} $\Phi$) and 2) the ``context'' of other vertexes (\textit{i.e.} $\Phi'$), so that the vertexes sharing similar context information can receive closed representations.

Consequently, minimizing the gap between the given distribution $p$ and the estimated distribution $\hat{p}$ helps preserve the similarity property in the representations:
\begin{equation}
    O = d(p(v_{j}|v_{i}), \hat{p}(v_{j}|v_{i}))
\label{eq:objective}
\end{equation}
Moreover, minimizing the gap by Kullback-Leibler (KL) divergence achieves the process of network embedding methods as long as sampling corresponding positive vertex pairs $S_{pos}$ and negative vertex pairs $S_{neg}$:
\begin{equation}
O = - \sum_{(v_{i},v_{j})\in S_{pos}} \log \hat{p}(v_{j}|v_{i})
 + \sum_{(v_{i},v_{k})\in S_{neg}} \log \hat{p}(v_{k}|v_{i}).
\label{eq:obj}
\end{equation}

\begin{algorithm}[htb]
\caption{Proposed Sampling-based Learning Framework}
\label{alg:framework}
\DontPrintSemicolon
\KwIn{A network graph $G$ with edges $E$ and weights $W$: $G(V, E, W)$}
Initialize the representations $\Phi$ and $\Phi'$ for $V$\;
\While{$\text{Conditions hold}$}{
    $S_{pos} \leftarrow \text{Sample positive vertex pair(s) according to $W$}$ \;
    $S_{neg} \leftarrow \text{Sample negative vertex pair(s) according to $W$}$ \;
    $\text{Update $\Phi$ and $\Phi'$ by minimizing Equation (\ref{eq:obj})} $
}
\KwOut{ Vertex representations via $\Phi$ }
\end{algorithm}

In the proposed framework, the positive pairs $S_{pos}$ and the negative pairs $S_{neg}$ are sampled according to the distribution of weights. Therefore, the framework can also be used to implement ``weighted'' variants of existing network embedding methods, such as DeepWalk. Algorithm~\ref{alg:framework} gives a pseudo code of the proposed weighted network embedding framework.  We can see that, by modifying the weights distribution of input network, the strength of the relations can be easily embedded in the representations.

\subsection{Asynchronous Optimization}

The optimization calculates the partial derivatives of the objective function by $\frac{\partial O}{\partial\Phi}$ and $\frac{\partial O}{\partial\Phi'}$ and updates the representation by the gradient direction like $\Phi = \Phi - \alpha \times \frac{\partial O}{\partial\Phi}$ and $\Phi' =
\Phi' - \alpha \times \frac{\partial O}{\partial\Phi'}$, where $\alpha$ serves as the learning rate.  The gradients of the log likelihood functions can be computed as:
\begin{equation}
  \frac{\partial log(\hat{p}(v_{i}|v_{j}))}{\partial \Phi(v_{i})}
  = (1-\sigma(\Phi(v_{i}) \cdot \Phi'(v_{j}))) \cdot \Phi'(v_{j})
\end{equation}
\begin{equation}
  \frac{\partial log(\hat{p}(v_{i}|v_{j}))}{\partial \Phi'(v_{j})}
  = (1-\sigma(\Phi(v_{i}) \cdot \Phi'(v_{j}))) \cdot \Phi(v_{i})
\end{equation}
We further apply Hogwild!~\cite{ASGD} for updating these representation values. This method is also known as the asynchronous stochastic gradient descent (ASGD) that updates the representations without any locking.  This makes the model update more efficient.

\section{Existing Non-Uniform Sampling Techniques}
\label{sec_related}

To sample the positive and negative vertex pairs $S_{pos}$ and $S_{neg}$ in Equation~(\ref{eq:obj}), we need to know which vertexes are in the proximities of each vertex.  
There are two kinds of pair searching strategy to realize this: 1) pre-calculate all the pair-wise vertex-vertex transition probabilities~\cite{grarep,dngr} and 2) dynamically collect the vertexes by random walks~\cite{dw,line,hpe,n2v}. 
This first method may not be efficient for a large information network. Therefore, in this paper we focus on the second strategy (i.e. using random walks), due to the relatively lower memory usage and better computational efficiency.


Many existing random walk-based network embedding methods apply the uniform sampling to retrieve vertex pairs from a given network, because it is easy to implement and it is efficient. However, this comes at the cost of ignoring the relation strength between two vertexes. A weighted sampling method that considers the edge weights may better capture the relationship among the vertexes. 

\begin{figure}
\centering
\includegraphics[width=0.45\textwidth]{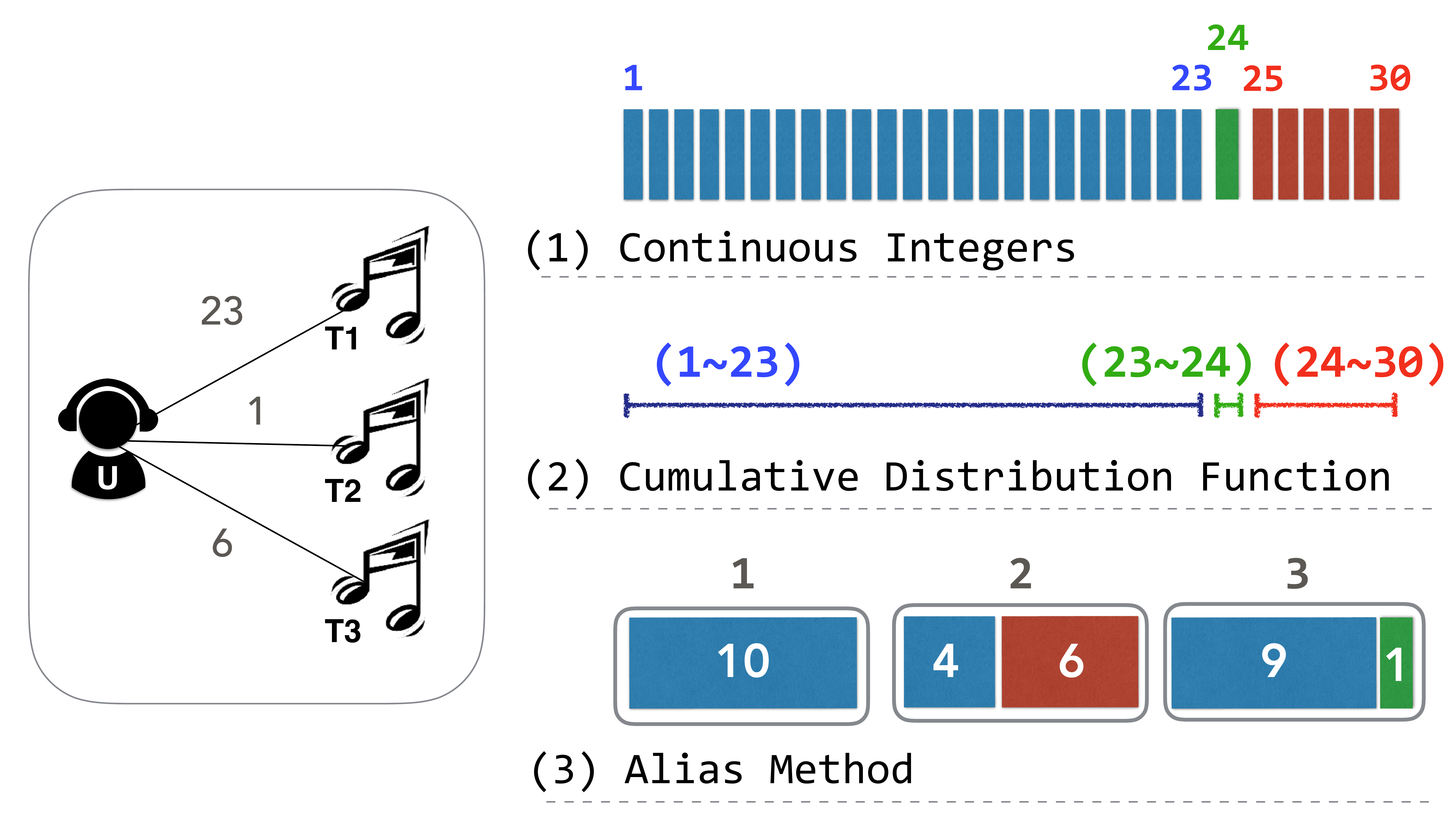}
\caption{Three different approaches to sampling a vertex from a non-uniform distribution.}
\label{fig:alias}
\end{figure}

According to our survey, there are three popular methods for drawing a sample from a non-uniform distribution. Suppose we plan to sample a music track from the total $n$ music tracks with corresponding listening frequency, Figure~\ref{fig:alias} shows the three different implementations.
\begin{enumerate}
  \item \textbf{Continuous Integers}\label{ht}: Set up a continuous integers distribution with size equal to the total number of possible edge weights. This forms a continuous discrete distribution, so sampling a random integer number in the range $[0, n]$ can be mapped into a certain vertex and follow the given frequency distribution.  This sampling method is the most efficient solution. Sampling a random integer number costs only $\mathcal{O}(1)$ time. A downside is that it can deal with only integer weights. As to the non-integer weights, it is needed to rescale them to a set of larger non-integer weights and thus occupies more than $\mathcal{O}(|V|)$ space, where $V$ is the set of all vertexes in the network.

  \item \textbf{Cumulative Density Function} (CDF)\label{cdf}: Compute the CDF and store the occupied range of each vertex, then to randomly sample a real number from $[0, sum(weights)]$ and check which range the number falls in. This sampling way is the most popular solution because it is applicable to the non-integer weights and occupies exact $\mathcal{O}(|V|)$ space. A drawback is that it is not a hashing method so it relies an additional search for finding out which cell the random number falls into. Using the binary search strategy requires $\mathcal{O}(log(|V|))$ time.

  \item \textbf{Alias Method}\label{al}~\cite{alias}: This is a technique that unfolds and reorganizes $n$ given weights into $n$ \textit{binary} cells with the equal size. This forms a continuous discrete distribution as well and requires only $n$ cells. By sampling a random real number from the range $[0, n)$, the integer part can be mapped to a certain bin consist of two vertexes and the decimal part help determine the which vertex to be received. This sampling technique costs only $\mathcal{O}(1)$ time for drawing a vertex.  This method uses one table for storing the alias probability and one table for storing the alias position, so the space complexity is $\mathcal{O}(|V|)$. 
\end{enumerate}



\begin{figure*}
  \begin{subfigure}{.45\textwidth}
    \centering
    \vspace{0.5cm}
    \includegraphics[width=.95\linewidth]{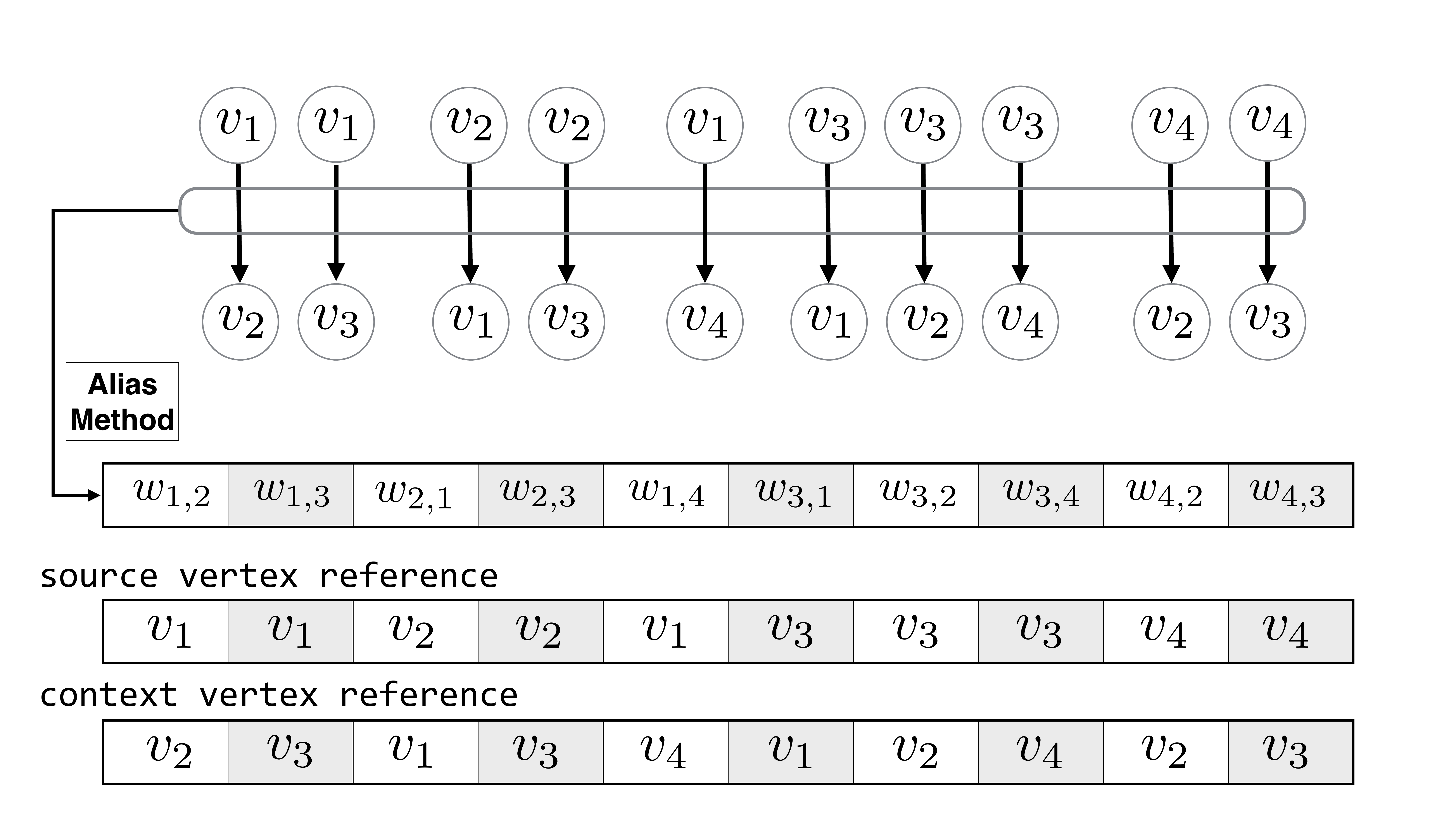}
    \vspace{0.5cm}
    \caption{The edges-based sampling method in LINE~\cite{line}.}
    \label{fig:edges}
  \end{subfigure}%
  \begin{subfigure}{.55\textwidth}
    \centering
    \includegraphics[width=.95\linewidth]{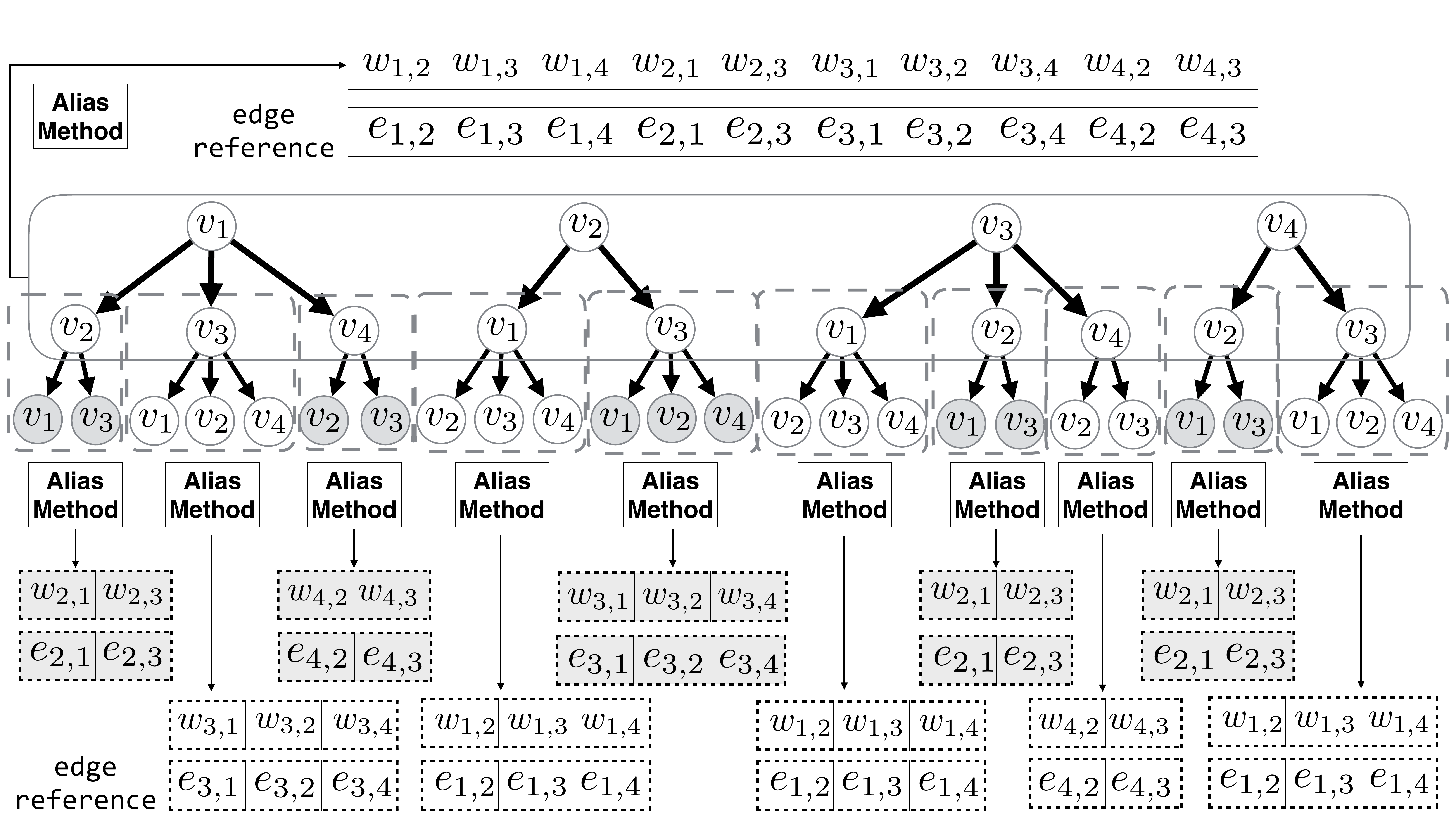}
    \caption{The edge-edge sampling method of node2vec~\cite{n2v}.}
    \label{fig:e2e}
  \end{subfigure}
  \caption{The two different sampling methods used in previous work.}
  \label{fig:method}
\end{figure*}

Among the preceding three solutions, the alias method is the only one that can keep both the time and space complexity low. In practice, however, the alias method is not applicable to an arbitrary data structure because the method reads a fixed distribution and needs a pre-computed cache for storing the mappings. Below we briefly review how two previous work, node2vec~\cite{n2v} and LINE~\cite{line}, use the alias method for weighted network embedding. 

\subsection{Edges Sampling Method}\label{sec:edge}

Figure~\ref{fig:edges} shows the data structure used in LINE~\cite{line} The edges graph network stores the whole distribution of the edges in a network, and maintains two tables that store the vertex reference (\textit{i.e.} source vertex) and the connected vertex (\textit{i.e.} context vertex). This data structure stores the edges in a simple and clear way. It greatly reduces the space usage and so LINE model is applicable to a large-scale network. One major concern is that it losses the ability to generate a walk to get the high-order proximities of a vertex, because there is no information about how the edges are connected to each other. In addition, the utilized edge-edge sampling strategy is used only for retrieving the positive connections, so it additionally claims a continuous integers distribution with for example $N=100$ million size for retrieving the negative pairs. The final demanded space complexity consist of:
\begin{itemize}
  \item The alias method for the edges: $\mathcal{O}(|E|)$.
  \item The two vertex reference tables: $\mathcal{O}(|E|)$.
  \item The continuous integers distribution: $\mathcal{O}(N)$.
\end{itemize}

\subsection{Edge-Edge Sampling Method}\label{sec:ee}

Figure~\ref{fig:e2e} shows the data structure used in node2vec~\cite{n2v}. The edge-edge sampling method stores all the connecting of edges, so it is possible to generate a random walk following a given edge by simply sampling the next edge one by one. This sampling method provides an advanced application usage in which we can re-assign the distribution of the connecting edges. The node2vec model proposed to modify the weights of the distribution based on the connection status of the two previous vertexes, such that increasing the weight when a vertex is connected to both the two vertexes, which makes it possible for breath-first search (BFS) or depth-first search (DFS) based random walks. However, the space complexity is huge.  The alias method is applied to the whole edge distribution (\textit{i.e.} the solid cycle in Figure~\ref{fig:e2e}) and the connected edges of every edge (\textit{i.e.} the dash circles in Figure~\ref{fig:e2e}).
The corresponding space complexity are $\mathcal{O}(|E|)$ and $\mathcal{O}(|E|\times avg|E(v)|)$ respectively, where $|E|$ is the total number of edges and $avg|E(v)|$ represents the average number of the connected edges of a given edge in the network.  Similarly, it additionally claims a continuous integers distribution with $N=100$ million size for retrieving the negative pairs. Please note that the alias method helps determine the cell to be selected, so this method also needs to store an edge reference to represent the specific edge it selects.  As an edge is composed of the source and the context vertexes, the overall space complexity becomes:
\begin{itemize}
  \item The alias table for edges: $\mathcal{O}(|E|+|E|\times avg|E(v)|)$.
  \item The edge reference table: $\mathcal{O}(|E|+|E|\times avg|E(v)|)$.
  \item The continuous integers distribution: $\mathcal{O}(N)$.
\end{itemize}

In short, due to the specific needs, LINE stores only the edge connections so it can sample only the single-step directly connected vertex of any vertex. Accordingly, it cannot exploit high-order proximities of any vertex. In contrast, node2vec searches the high-order proximities by BFS and SFS, so it proposes to stores all the edge-edge connections. This demands a great amount of memory space for a large network. 

To address the aforementioned two issues simultaneously, we describe in the next section a novel vertex-context sampling method that applies the alias method to a delicate sequential vertex-context graph network, which is capable of searching the high-order proximities of a vertex with low time and space complexity.





\section{Proposed Data Structure and Sampling Method}\label{sec_proposed}

\subsection{Vertex-Context Sampling Method}\label{sec:vc}

\begin{figure}
  \centering
  \includegraphics[width=0.45\textwidth]{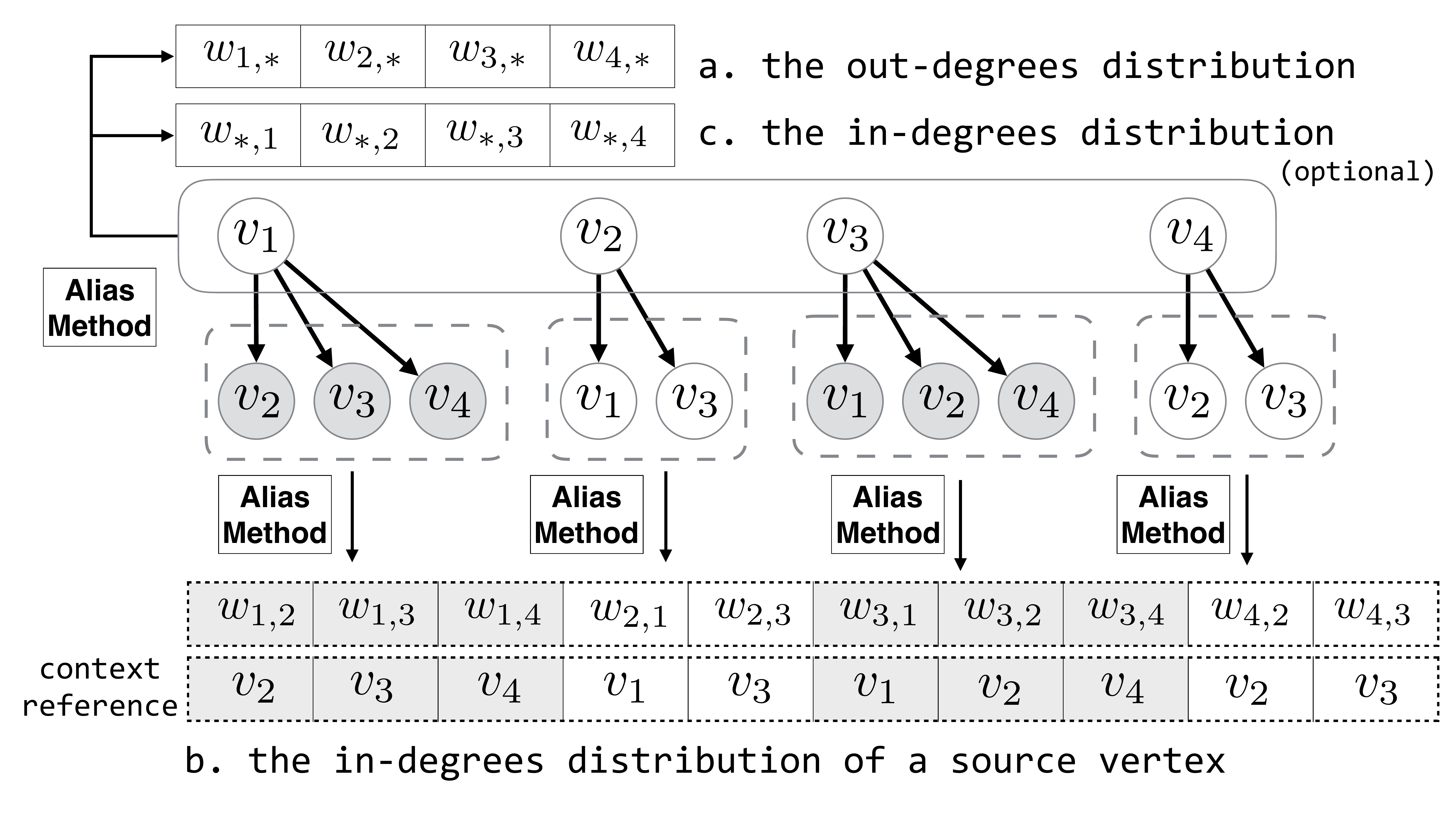}
  \caption{Sequential vertex-context graph data structure for generalized weighted network embedding.}
  \label{fig:v2c}
\end{figure}

Figure~\ref{fig:v2c} shows the data structure employed by the proposed method. It stores the out-degrees distribution and in-degrees distribution separately. For the out-degrees distribution (\textit{i.e} Figure~\ref{fig:v2c}a), as the weights are stored in a table in ascending order, the index of a cell is the same as the index of a vertex.  Hence, there is no need to maintain an additional table for storing the vertex reference and thus save the memory space.  For instance, the second cell of the alias table represents the second vertex in the network. For the in-degrees distribution (\textit{i.e} Figure~\ref{fig:v2c}b), the alias method can be applied to the sub-blocks (\textit{i.e.} the dash circles) separately, so each sub-block stores the in-degrees distribution of a certain vertex.  Sampling a vertex in a sub-block is equivalent to sampling a context (a connected vertex) of a vertex. Suppose that there are $|V|$ vertexes in the given network, the overall space complexity consists of:
\begin{itemize}
  \item The alias table for source vertexes: $\mathcal{O}(|V|)$.
  \item The alias table for context vertexes: $\mathcal{O}(|E|)$.
  \item The vertex reference for context vertexes: $\mathcal{O}(|E|)$.
\end{itemize}
Note that many existing models, including DeepWalk, nod2vec, and LINE, apply the continuous integers distribution to realize negative sampling. In contrast, with this data structure we can sample negative pairs via either the context in-degrees distribution (\textit{i.e} Figure~\ref{fig:v2c}b) or the vertex in-degrees distribution (\textit{i.e} Figure~\ref{fig:v2c}c), which will be small when $|V| < 100,000,000$.

\begin{figure}
  \centering
  \includegraphics[width=0.45\textwidth]{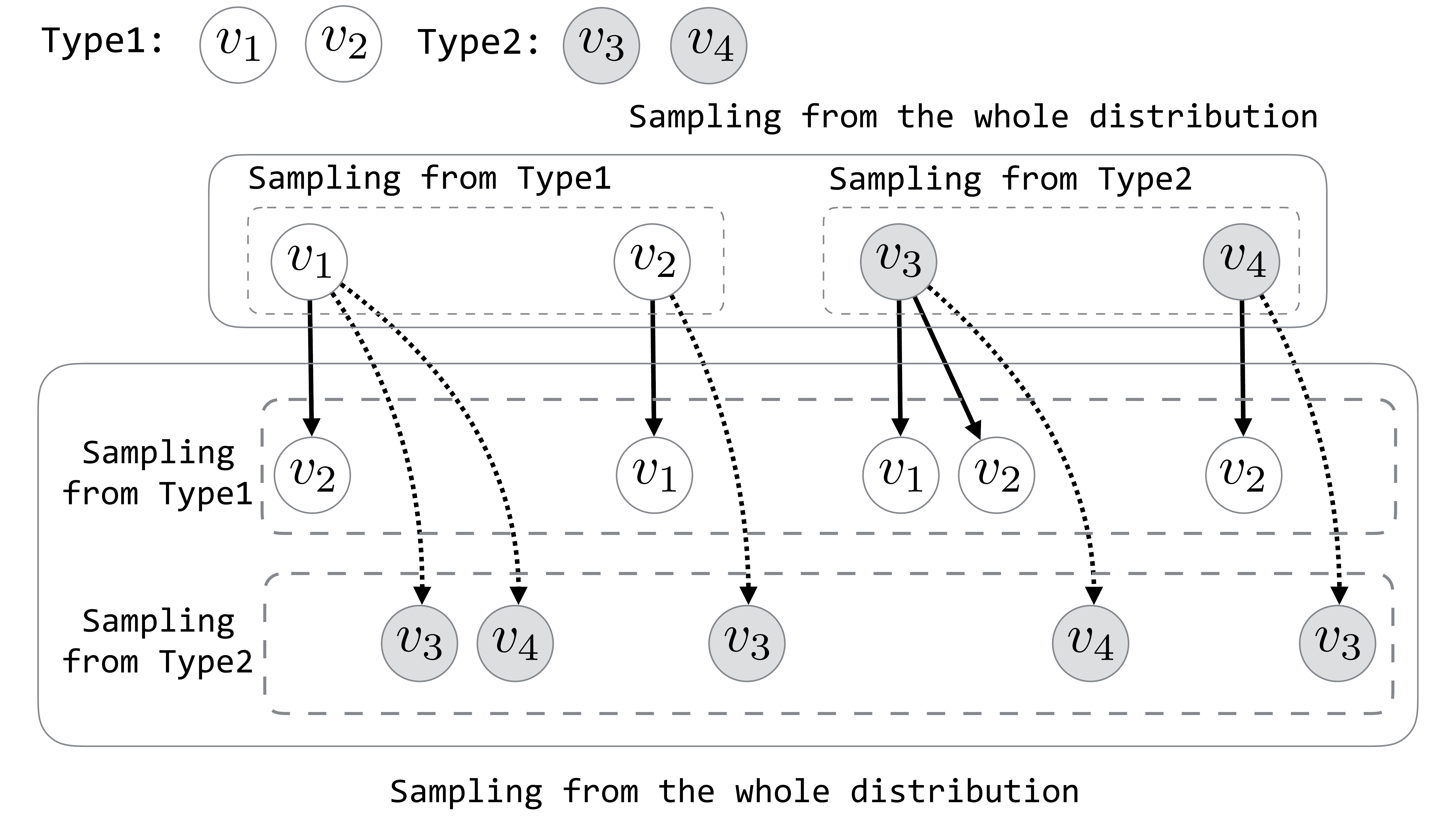}
  \caption{An example for type-aware vertex-context sampling.}
  \label{fig:v2cb}
\end{figure}

Besides, it can be observed that concatenating all the sub-blocks is equivalent to storing the edges distribution.  This property is useful when we need to sample a vertex or a context by certain characteristic.  Specifically, the sequential graph network can store multiple context distributions, so it offers the flexibility of sampling pairs within certain group.  Figure~\ref{fig:v2cb} depicts an example network which consists of two types of vertexes. By storing the vertex distributions and context distributions with their types, the data structure permits sampling a vertex or a context according to either a single type or the whole distribution.

We can implement various existing network embedding methods based on the proposed framework. In doing so, we produce three essential kinds of sampling functions by the proposed network data structure:
\begin{enumerate}
  \item $VertexSampling$: Sample a vertex according to the whole out-degrees distribution.
  \item $ContextSampling(v)$: Sample a vertex according to the in-degrees distribution of a source vertex $v$.
  \item $NegativeSampling$: Sample a vertex according to the whole in-degree distributions. In practice, the weights of negative sampling are transformed by a logarithm function.
\end{enumerate}
Consequently, searching the higher-oder proximities can still be done by generating a random walk over the graph.  It is equivalent to sampling a vertex according to the out-degrees in the first and then to iteratively sample the context of previous vertex, which forms the idea of the vertex-context sampling method.





\begin{algorithm}[htb]
\caption{VCS-DeepWalk}
\label{alg:dw}
\DontPrintSemicolon
\KwIn{ Network Graph with Weights: $G(V, E, W)$, Learning Rate: $\alpha$
Walk Times: $t$, Walk Length: $l$, Window Size: $w$, Negative Samples: $e$
}

\For{ $v \in V$ }{
Initialize the representations: $\Phi(v)$, $\Phi'(v)$ \;
}

\For{ $i \in \{1, ..., t\}$ }{
    \For{ $v \in \text{shuffle}(V)$ }{
        $Walk = \{v\}$ \;
        \For{ $i \in \{1, ..., l\}$ }{
            $v_{pos} = ContextSampling(Walk[-1])$ \;
            $Walk = Walk \cup v_{pos}$ \;
        }
        \For{ $v_{i} \in Walk$ }{
            \For{ $v_{j} \in Walk[i-w:i+w]$ }{
                $S_{pos} = \{(v_{i}, v_{j})\}$ \;
                $S_{neg} = \{\}$ \;
                \For { $v' \in  \{1, ..., e\}$ }{
                    $v_{neg} = NegativeSampling()$ \;
                    $S_{neg} = S_{neg} \cup (v_{i}, v_{neg})$ \;
                }
                Update $\Phi$, $\Phi'$ by minimizing Eq.~(\ref{eq:obj}) \;
            }
        }
    }
}

\KwOut{ Vertex representations $\Phi$ }
\end{algorithm}

\section{Weighted Network Embedding}
\label{sec:wne}

In this section, we demonstrate how to exploit the developed vertex-context sampling functions to implement various weighted network embedding methods, and provide the corresponding pseudo algorithms.  In the experiments, we target on refining the generalized network embedding methods, including DeepWalk~\cite{dw}, Walklets~\cite{wl}, LINE~\cite{line} and HPE~\cite{hpe}.
As our solution will iteratively sample the vertex and the corresponding contexts, we named the design process as the Vertex-to-Context Sampling (VCS) method.


\subsection{VCS-DeepWalk}

Algorithm~\ref{alg:dw} is the implementation of DeepWalk~\cite{dw} model. The original DeepWalk algorithm generates the random walks on every vertex and uses a slicing window (\textit{i.e.} line 10) to circle the demanded positive vertex-context pairs for $S_{pos}$.  In a similar way, we propose to use $VertexSampling$ function to obtain a starting vertex $v$ for a walk, and then to iteratively apply $ContextSampling(v)$ function to sample a following connected vertex according to the edges distribution of the previous given vertex $v$. In Algorithm~\ref{alg:dw}, the walk generated in lines 5-8 is the present weighted random walk which additionally considers the edge relation strength. Lines 10-11 extracts the positive vertex-context pairs from a given window size and lines 12-15 produces the negative vertex-context pairs.

\subsection{VCS-Walklets}

Walklets~\cite{wl} is an extended version of DeepWalk model.  The key change is to model the $l$-step vertex-context pairs separately.  Concretely, they divide the $S_{pos}$ into $l$ individual collections: $\{S_{pos}^{1}$, $S_{pos}^{2}$, ..., $S_{pos}^{l}\}$, and then to model the representation in each set.  In this case, we can also replace the line 10 by considering only a specified window position like $v_{j} \in [W_{v}[i-w],W_{v}[i-+w]]$ to stimulate the Walklets process.  The final obtained representation is the concatenation of the representation in every set.  It is reported that the vertex-context pairs in $S_{pos}^{2}$ and $S_{pos}^{3}$ provide the most useful information of a vertex in the conducted experiments.

\subsection{VCS-LINE}

Algorithm~\ref{alg:line} is the implementation of LINE~\cite{line} model.  The original claims are to model two types of the connections. One is the first-order connections.  The goal is to map the connected two vertexes with similar representations. The other one is the second-order connections.  The goal is to map the vertexes which have similar connections with similar representations. By modeling the two types, it is shown to receive a better performance in variant prediction tasks.  Please note that LINE is already a weighted network embedding method, and we provide an alternative implementation though the developed sampling functions in this work, as shown in lines 5-6 and 14-15. Besides, VCS-LINE is feasible to get high-order relations once sampling the walks from the given network.

\begin{algorithm}[htb]
\caption{VCS-LINE}
\label{alg:line}
\DontPrintSemicolon
\KwIn{ Network Graph with Weights: $G(V, E, W)$, Sample Times: $t$,
Negative Samples: $e$}

\For{ $v \in V$ }{
Initialize the representations: $\Phi_{o1}(v)$, $\Phi_{o2}(v)$, $\Phi_{o2}^{'}(v)$ \;
}

\For{ $i \in \{1, ..., t\}$ }{
    
    $v_{1} = VertexSampling()$ \;
    $v_{2} = ContextSampling(v_{1})$ \;
    $S_{pos} = \{ (v_{1}, v_{2}) \}$ \;
    $S_{neg} = \{\}$ \;
    \For { $v' \in  \{1, ..., e\}$ }{
        $v_{neg} = NegativeSampling()$ \;
        $S_{neg} = S_{neg} \cup (v_{1}, v_{neg})$ \;
    }
    Update $\Phi_{o1}$ by minimizing Eq.~(\ref{eq:obj}) \;
    Update $\Phi_{o2}$, $\Phi_{o2}^{'}$ by minimizing Eq.~(\ref{eq:obj}) \; 
}

\KwOut{ Vertex representations $\Phi_{o1}$ and $\Phi_{o2}$ }
\end{algorithm}

\begin{algorithm}[htb]
\caption{VCS-HPE}
\label{alg:hpe}
\DontPrintSemicolon
\KwIn{ Network Graph with Weights: $G(V, E, W)$, Sample Times: $t$,
Walk Length: $l$, Negative Samples: $e$}

\For{ $v \in V$ }{
Initialize the representations: $\Phi(v)$, $\Phi'(v)$ \;
}

\For{ $i \in \{1, ..., t\}$ }{
    $v_{1} = VertexSampling()$ \;
    $v_{next} = ContextSampling(v_{1})$ \;

    \For{ $i \in \{1, ..., l\}$ }{
        $S_{pos} = \{(v_{1}, v_{next})\}$ \;
        $S_{neg} = \{\}$ \;
        \For { $v' \in  \{1, ..., e\}$ }{
            $v_{neg} = NegativeSampling()$ \;
            $S_{neg} = S_{neg} \cup (v_{i}, v_{neg})$ \;
        }
        Update $\Phi$, $\Phi'$ by minimizing Eq.~(\ref{eq:obj}) \;
        $v_{next} = ContextSampling(v_{next})$ \;
    }
}

\KwOut{ Vertex representations $\Phi$ }
\end{algorithm}

\subsection{VCS-HPE}

Algorithm~\ref{alg:hpe} is the implementation of HPE~\cite{hpe}.  It is a fusion approach of DeepWalk and LINE.  The retrieved contexts are obtained from a short random walk and the updating frequency of a vertex follows the edges distribution.  One novel change is that it updates only the representations of the starting vertex, which makes the loss function following the exact out-degrees distribution.


\subsection{Discussion}
The node2vec simulation is not reported in the conducted experiments because of the three reasons below. 1) Although it is able to use vertex-context sampling to generate BFS/DFS walk, the vertex-based sampling approach is still not identical to the original mechanism. Please refer to \cite{n2v} for details. However, the proposed framework is memory efficient. 2) node2vec is already a weighted network embedding model. 3) The original release of node2vec is not feasible to the large-scale networks presented in our experiments.

In addition to network embedding, the proposed framework is also applicable to many other representation learning techniques, such as word2vec~\cite{w2v}, GloVe~\cite{glove}, matrix factorization (MF), factorization machine (FM)~\cite{fm}, etc. For instance, the sentences of an document can be converted to a word-to-word network, and the rating matrix can be converted to a user-to-item network. For clear presentation in this paper, we leave them as the future work.

\section{Experiments}\label{sec:exp}

\subsection{Experimental Datasets}

We conduct the experiments on one public wikipedia dataset named \footnote{http://www.mattmahoney.net/dc/textdata}{text9}, and one public movie rating dataset named \footnote{https://grouplens.org/}{movielens-latest}. The wiki dataset is built as the language network that words within every 5-word sliding window are treated as the co-occurring word with each other. The words with frequency smaller than 5 are filtered out.  The movielens datasets was generated on October 18, 2016. It is built as the preference network that the edge weights are determined by the 5-star ratings.  

In addition, we also prepare an on-line evaluation for examining the performance difference with and without non-uniform weights.  The task is done by accessing the music streaming service on \footnote{The leading music streaming company is Asia.}{KKBOX} platform. The built preference network contains the number of listening times to the music for each user. Table~\ref{tb:data} summarizes the statistics of the three examined datasets.

\begin{table}%
\centering
\begin{tabular}{lrrc}
\toprule
Dataset & \#Vertex &  \#Edge & Network Type \\ 
\midrule
wiki-text9		    & 218,316	& 112,090,997	& Language \\
movielens-latest	& 299,247	& 24,404,096	& Preference \\
KKBOX (online)        & 3,000,000	& 660,000,000	& Preference \\
\bottomrule
\end{tabular}
\caption{Statistics of Experimental Datasets}
\label{tb:data}
\end{table}%

\subsection{Performance Evaluations}

For the language network, we examine the contextual word similarity which is a standard method for evaluating the word vector embeddings. There are \footnote{http://www.wordvectors.org/}{eight public benchmarks} used in our experiments.  Each dataset contains a set of word pairs and the corresponding similarity are judged by humans on a certain scale. The Spearman' correlation between the benchmarks and learned representations is reported in this task. It provides the insights into how well the learned word vectors capture the context information.

For the preference network, we perform two tasks.  One is the tag recommendations, as demonstrated in \cite{dw,line}.  We follow the same setting and aim at predicting the correct tags of a movie.  This supervised prediction task is predicted by training the one-vs-rest logistic regression classifier on the learned representations using the \footnote{http://www.csie.ntu.edu.tw/\~cjlin/liblinear/}{LIBLLINEAR} package. We report the classification metrics Micro-F1 and Macro-F1 in this task when uses 10\%-50\% of training labels.  The second one is the item-item recommendations like the tasks performed in \cite{hpe,next}.  This task is similar to the query-based cold start recommendations and the next-song recommendations.  We randomly select 80\% of the rating records for each user as the training data, and put the rest 20\% records in the test set for the off-line evaluation.  In the testing stage, we randomly select 5 movie queries for each user, and ask for matching the recommendations to the selected queries.  We report the score of Recall at $k$ (Recall@k), HitRatio at k (HR@k) and mean Average Precision at k (mAP@k).  All the results are averaged over 10 distinct runs by sampling different training and testing data.

\subsection{Different Weighting Schemes}

The proposed weighted network embedding methods can sample the training vertex pairs according to arbitrary edge weights. It changes the appearing frequency in $P_{pos}$ and $P_{neg}$ in Equation~\ref{eq:obj}. Hence, different weighting schemes help capture different views of the network structures.  In language networks, we present three weighting schemes:
\begin{enumerate}
  \item \textbf{Binary}: Convert the edge weights to be 0/1.  This method equals to the uniform network embedding methods.
  \item \textbf{TF}: Use the term frequency as the edge weights.  High frequency means the two words are considered to be much more similar to each other.
  \item \textbf{TF-IDF}: Use the term-frequency-inverse-document-frequency (TF-IDF) as the edge weights.  The IDF highlights the important rare words.
\end{enumerate}
In preference network, we present the same three weighting schemes:
\begin{enumerate}
  \item \textbf{Binary}: Convert the edge weights to be 0/1.  This method equals to the uniform network embedding methods.
  \item \textbf{Rating}: Use the ratings as the edge weights.  High rating score means the user favors the movie more.
  \item \textbf{Rating-IRF}: Make use of the concept of the IDF, by multiplying the rating with the inverse rating frequency (IRF): $IRF_{i} = log\frac{|U|}{|\{u:r_{u,i}\neq 0\}|}$. The idea is that the rarely rated movies may contribute more information than the frequently rated movies.
\end{enumerate}

\subsection{Experimental Results}

{
\begin{table*}%
\centering
\begin{tabular}{lcccccccc}
\toprule
 & {\small WS-353-ALL} & {\small MTurk-771} & {\small VERB-143}  & {\small MEN-TR-3k} & {\small YP-130} & {\small SIMLEX-999} & {\small WS353-REL} & {\small RW-Stanford} \\
\midrule
LINE~\cite{line} [Binary]   & 62.56\% & 55.79\% & 34.83\% & 63.88\% & 40.28\% & 25.52\% & 51.07\% & 26.96\% \\
VCS-LINE [Binary]           & 65.14\% & 56.06\% & 32.95\% & 64.88\% & 37.77\% & 26.24\% & 54.43\% & 32.36\% \\
\midrule
word2vec \cite{w2v}     & 72.54\% & 65.91\% & 40.07\% & 74.64\% & 44.80\% & 35.98\% & 78.05\% & 37.72\% \\
VCS-DeepWalk [Binary]   & 69.46\% & 59.47\% & 32.44\% & 71.71\% & 48.17\% & 30.90\% & 68.92\% & 36.57\% \\
VCS-DeepWalk [TF]       & 72.18\% & 64.69\% & 38.10\% & 74.41\% & 49.03\% & 33.61\% & 72.73\% & 37.98\% \\
VCS-DeepWalk [TF-IDF]   & 72.69\% & 63.73\% & 32.54\% & 74.08\% & 49.79\% & 32.96\% & 71.28\% & 37.05\% \\
\midrule
VCS-Walklets [Binary]   & 54.24\% & 50.31\% & 32.09\% & 63.29\% & 29.53\% & 19.20\% & 51.90\% & 25.10\% \\
VCS-Walklets [TF]       & 58.65\% & 51.58\% & 29.88\% & 65.39\% & 34.02\% & 20.17\% & 63.83\% & 27.27\% \\
VCS-Walklets [TF-IDF]   & 59.24\% & 53.31\% & 28.33\% & 67.00\% & 28.12\% & 21.70\% & 57.06\% & 28.56\% \\
\midrule
VCS-HPE [Binary]        & 73.59\% & 63.87\% & 32.57\% & 73.43\% & 44.69\% & 34.38\% & 73.17\% & 37.79\% \\
VCS-HPE [TF]            & 72.78\% & 66.75\% & 36.57\% & 75.39\% & 47.43\% & 36.52\% & 74.06\% & 40.33\% \\
VCS-HPE [TF-IDF]        & 74.30\% & 65.13\% & 32.99\% & 75.23\% & 48.22\% & 34.36\% & 75.18\% & 39.00\% \\
\bottomrule
\end{tabular}
\caption{Results of word similarity search on the text9 language network}
\label{tb:text9_word_sim}
\end{table*}%
}

{

\begin{table*}%
\centering
\begin{tabular}{lcccccccccc}
\toprule
Train Ratios & \multicolumn{2}{c}{10\%} & \multicolumn{2}{c}{20\%} & \multicolumn{2}{c}{30\%} & \multicolumn{2}{c}{40\%} & \multicolumn{2}{c}{50\%} \\
\midrule
Metrics & {\small micro-f1} & {\small macro-f1} & {\small micro-f1} & {\small macro-f1} & {\small micro-f1} & {\small macro-f1} & {\small micro-f1} & {\small macro-f1} & {\small micro-f1} & {\small macro-f1}\\
\midrule
LINE \cite{line} [Binary]   & 53.73\% & 43.11\% & 55.20\% & 45.34\% & 55.76\% & 46.11\% & 56.24\% & 47.07\% & 56.50\% & 47.48\% \\
VCS-LINE [Binary]           & 53.67\% & 43.09\% & 54.88\% & 44.92\% & 55.52\% & 45.90\% & 56.03\% & 47.00\% & 56.20\% & 47.13\% \\
\midrule
DeepWalk \cite{dw}          & 49.71\% & 39.13\% & 50.55\% & 41.40\% & 50.80\% & 42.04\% & 51.04\% & 42.52\% & 51.18\% & 42.80\% \\
VCS-DeepWalk [Binary]       & 49.00\% & 36.97\% & 50.19\% & 40.13\% & 50.78\% & 41.22\% & 51.15\% & 41.71\% & 51.34\% & 42.19\% \\
VCS-DeepWalk [Rating]       & 48.56\% & 36.86\% & 49.84\% & 39.81\% & 50.23\% & 40.47\% & 50.60\% & 41.09\% & 50.89\% & 41.59\% \\
VCS-DeepWalk [Rating-IRF]   & \textbf{51.21}\% & \textbf{39.76}\% & \textbf{52.20}\% & \textbf{42.83}\% & \textbf{52.59}\% & \textbf{43.72}\% & \textbf{52.97}\% & \textbf{44.43}\% & \textbf{53.10}\% & \textbf{44.51}\% \\
\midrule
VCS-Walklets [Binary]       & 43.66\% & 30.80\% & 44.61\% & 33.67\% & 45.01\% & 34.54\% & 45.27\% & 34.96\% & 45.38\% & 35.31\% \\
VCS-Walklets [Rating]       & 43.71\% & 31.05\% & 44.44\% & 33.53\% & 44.57\% & 34.39\% & 45.07\% & 35.16\% & 45.34\% & 35.47\% \\
VCS-Walklets [Rating-IRF]   & \textbf{47.07}\% & \textbf{34.23}\% & \textbf{47.94}\% & \textbf{36.99}\% & \textbf{48.17}\% & \textbf{37.77}\% & \textbf{48.48}\% & \textbf{38.52}\% & \textbf{48.79}\% & \textbf{39.00}\% \\
\midrule
VCS-HPE [Binary]        & 52.72\% & 42.67\% & 53.79\% & 45.28\% & 54.03\% & 46.14\% & 54.37\% & 46.79\% & 54.55\% & 46.91\% \\
VCS-HPE [Rating]        & 52.01\% & 42.14\% & 53.05\% & 44.77\% & 53.47\% & 45.43\% & 53.78\% & 45.92\% & 53.74\% & 46.16\% \\
VCS-HPE [Rating-IRF]    & \textbf{53.66}\% & \textbf{44.02}\% & \textbf{54.65}\% & \textbf{46.63}\% & \textbf{55.09}\% & \textbf{47.25}\% & \textbf{55.31}\% & \textbf{47.77}\% & \textbf{55.30}\% & \textbf{47.77}\% \\
\bottomrule
\end{tabular}
\caption{Results of genre tag recommendations on the MovieLens-latest preference network}
\label{tb:ml-latest_tag}
\end{table*}%
}

{

\begin{table*}%
\centering
\begin{tabular}{lccccccccc}
\toprule
Metrics                        & {\small Recall@10} & {\small HR@10} & {\small mAP@10} & {\small Recall@20} & {\small HR@20} & {\small mAP@20} & {\small Recall@30} & {\small HR@30} & {\small mAP@30} \\
\midrule
LINE \cite{line} [Binary]   & 12.76\% & 12.36\% & 7.04\% & 12.17\% & 10.82\% & 5.31\% & 12.22\% & 9.92\% & 4.61\%  \\
VCS-LINE [Binary]           & 12.84\% & 12.26\% & 7.03\% & 12.53\% & 10.86\% & 5.42\% & 12.79\% & 9.97\% & 4.79\%  \\
\midrule
DeepWalk \cite{dw}          & 11.33\% & 10.80\% & 6.33\% & 10.30\% & 8.84\% & 4.57\% & 10.05\% & 7.76\% & 3.87\%  \\
VCS-DeepWalk [Binary]       & 11.97\% & 11.40\% & 6.76\% & 10.58\% & 8.90\% & 4.72\% & 9.92\% & 7.41\% & 3.88\% \\
VCS-DeepWalk [Rating]       & 10.91\% & 10.37\% & 6.03\% & 9.42\% & 7.86\% & 4.14\% & 8.86\% & 6.56\% & 3.41\% \\
VCS-DeepWalk [Rating-IRF]   & \textbf{12.28}\% & \textbf{11.70}\% & \textbf{6.97}\% & \textbf{11.04}\% & \textbf{9.29}\% & \textbf{4.95}\% & \textbf{10.52}\% & \textbf{7.81}\% & \textbf{4.11}\% \\
\midrule
VCS-Walklets [Binary]       & 11.61\% & 10.98\% & 6.56\% & 10.31\% & 8.50\% & 4.65\% & 9.77\% & 7.11\% & 3.89\% \\
VCS-Walklets [Rating]       & 10.76\% & 10.16\% & 6.05\% & 9.51\% & 7.79\% & 4.25\% & 8.98\% & 6.47\% & 3.54\% \\
VCS-Walklets [Rating-IRF]   & \textbf{11.69}\% & \textbf{11.08}\% & \textbf{6.53}\% & \textbf{10.46}\% & \textbf{8.67}\% & \textbf{4.67}\% & \textbf{9.97}\% & \textbf{7.26}\% & \textbf{3.92}\% \\
\midrule
VCS-HPE [Binary]        & 11.03\% & 10.51\% & 6.08\% & 9.62\% & 8.14\% & 4.19\% & 9.01\% & 6.80\% & 3.44\% \\
VCS-HPE [Rating]        & 10.12\% & 9.67\% & 5.62\% & 8.79\% & 7.51\% & 3.81\% & 8.26\% & 6.27\% & 3.11\% \\
VCS-HPE [Rating-IRF]    & \textbf{11.67}\% & \textbf{11.11}\% & \textbf{6.59}\% & \textbf{10.53}\% & \textbf{8.91}\% & \textbf{4.62}\% & \textbf{10.01}\% & \textbf{7.51}\% & \textbf{3.83}\% \\
\bottomrule
\end{tabular}
\caption{Results of movie-movie recommendations on the MovieLens-latest preference network}
\label{tb:ml-latest_rec}
\end{table*}%
}

In the conducted experiments, we use VCS to represent the proposed weighted vertex sampling method.  Our examined proposed models include 1) VCS-LINE, 2) VCS-DeepWalk, 3) VCS-Walklets and 4) VCS-LINE.  In order to receive a fair comparison result, we limit the total of updating times by 2,000 millions and use the default model learning parameters.

Tables \ref{tb:text9_word_sim} shows the evaluated results of the word similarity measurements on eight benchmarks for the language networks.  Our VCS-LINE model receives a competitive performance with the original LINE model, and the VCS-DeepWalk receives a competitive performance with word2vec as well.  Please note that DeepWalk is an extended version of word2vec, but word2vec performs better in learning word representstions because it updates the representations following words distributions. This results confirms the validity of the developed algorithms. The exact performance difference may produced by the implementation details, such as the sub-sampling, reduced window size, etc.  In addition, the best performance of the benchmarks are almost all obtained by the non-uniform weighted network embedding methods.  These results support our claim that varying the edge weights could help better capture the locality (or context information) of a vertex in the network embedding methods.

Tables \ref{tb:ml-latest_tag} shows the evaluated results of the genre tag recommendations for the preference networks. Likewise, the VCS-LINE and the VCS-DeepWalk receive a competitive performance with the original versions.  Moreover, the TF-IDF weighting outperforms than others when use 10\%-50\% training data.  This provides an insight that the rarely rated movies may contribute more information during the optimization stage. Another interesting observation is that the direct use of the rating score may decrease the performance.  We consider this situation is caused by the dataset type and the examined task.  In genre tag recommendation task, the rated/unrated records provide more information about the user preference than the rating scores, which means the users tend to watch a movie in a preferred type but not purely based the user ratings.

Tables \ref{tb:ml-latest_rec} shows the evaluated results of the movie-movie recommendations for the same preference networks. DeepWalk earns a better performance at the cut-off position 20 and 30 on movielens-20m dataset.  We consider that, in this case, DeepWalk adopts the hierarchical softmax, an approximation of modeling multi-classifications, can achieve a better performance than the negative sampling method we adopted. They are two different solutions for modeling the representations with positive label only dataset.  Nevertheless, the developed weighted network embedding methods can still surpass DeepWalk in movielens-latest dataset.  In movie-movie recommendations, the rarely rated movies also contribute more information for a user's preference so that the Rating-IRF weighting methods perform the best.

\subsection{KKBOX On-line Recommendation Results}

The ground truth of the on-line testing process is determined by the user responses on-the-fly.  The training records are collected from 1-year listening history during 2016.  The recommendations for a user are generated based on the recent music played by the user, as the setting of the movie-movie recommendation task.  We applied three models, including 1) DeepWalk, 2) uniform VCS-HPE and 3) non-uniform VCS-HPE, to learn the representations.  The final obtained mAP of the DeepWalk, the uniform VCS-HPE and the non-uniform VCS-HPE are around 6.8\%, 8.2\% and 14.5\% respectively.

According to the results, we concluded that the first performance gap between the DeepWalk and the uniform VCS-HPE is caused by the edge sampling idea, which considers the edge weights distribution during the optimization stage.  The second performance gap between the uniform VCS-HPE and the non-uniform VCS-HPE can validate effectiveness of the weighted sampling approach.

\subsection{Memory Consumption}

\begin{figure}
\centering
\includegraphics[width=0.44\textwidth]{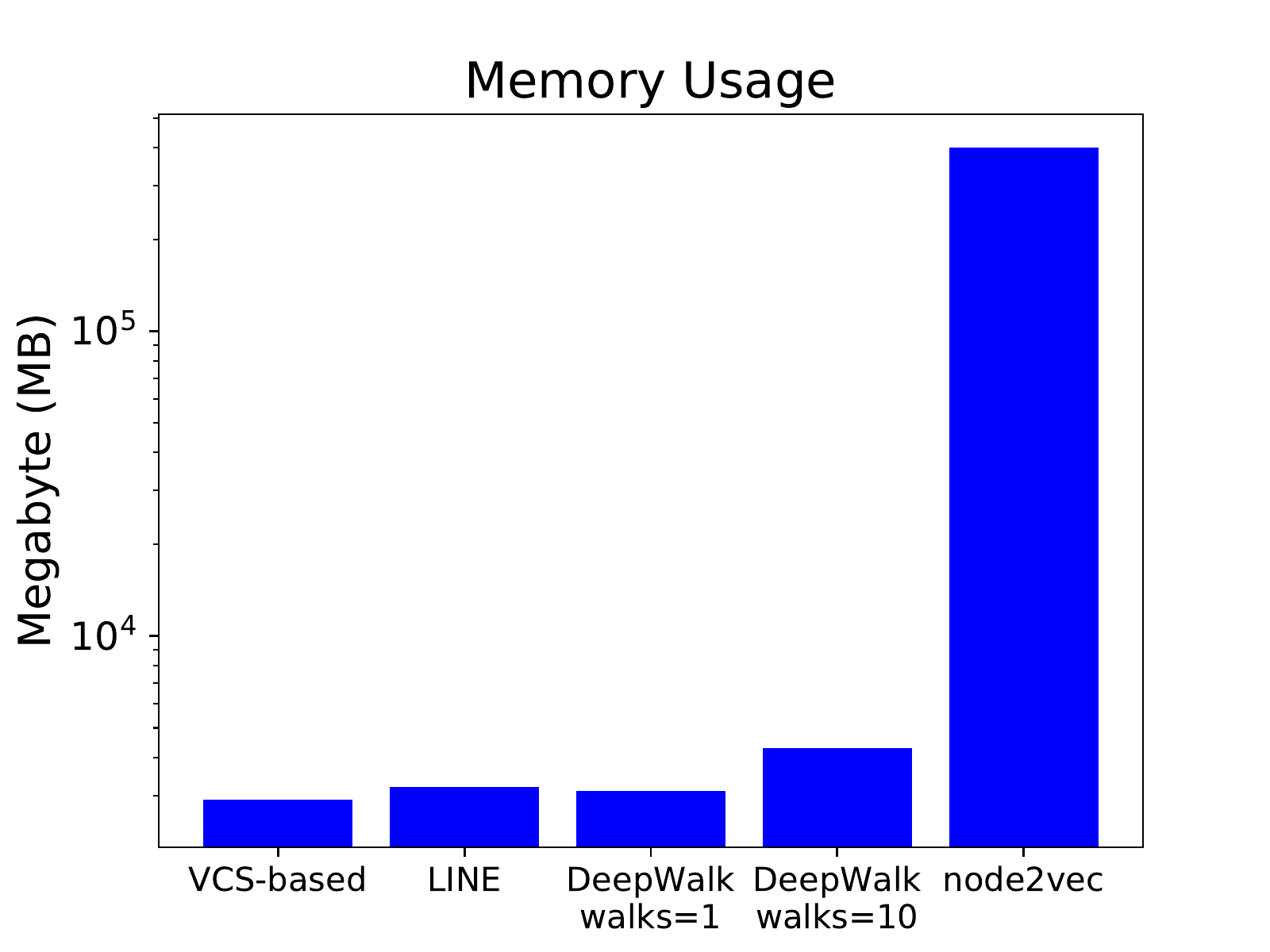}
\caption{Memory usages (MB) on movielens-latest dataset.}
\label{fig:memory}
\end{figure}


Figure~\ref{fig:memory} plots the memory usage of four network embedding methods on the movielens-latest dataset.  The occupied space of the proposed VCS-based network embedding approach is closed to LINE and DeepWalk (with single walk time).  The original LINE algorithm learns the 1st and 2nd representations separately (\textit{i.e.} using two execution programs), so we report the stacked results. The original DeepWalk model pre-computes the random walks so that the memory usage is increased when uses a large walk times. Note that it is possible to generate and train a specific pair at a time in our developed framework. The edge-edge graph structure used in node2vec is not feasible for a network containing a large number of edges.  It occupies more than 10 times that amount of space of the other models.  Besides, the examined DeepWalk and node2vec algorithms require a pre-process on mapping the objects to integers, whereas VCS-based and LINE do not.

\subsection{Execution Time}

\begin{figure}
\centering
\includegraphics[width=0.44\textwidth]{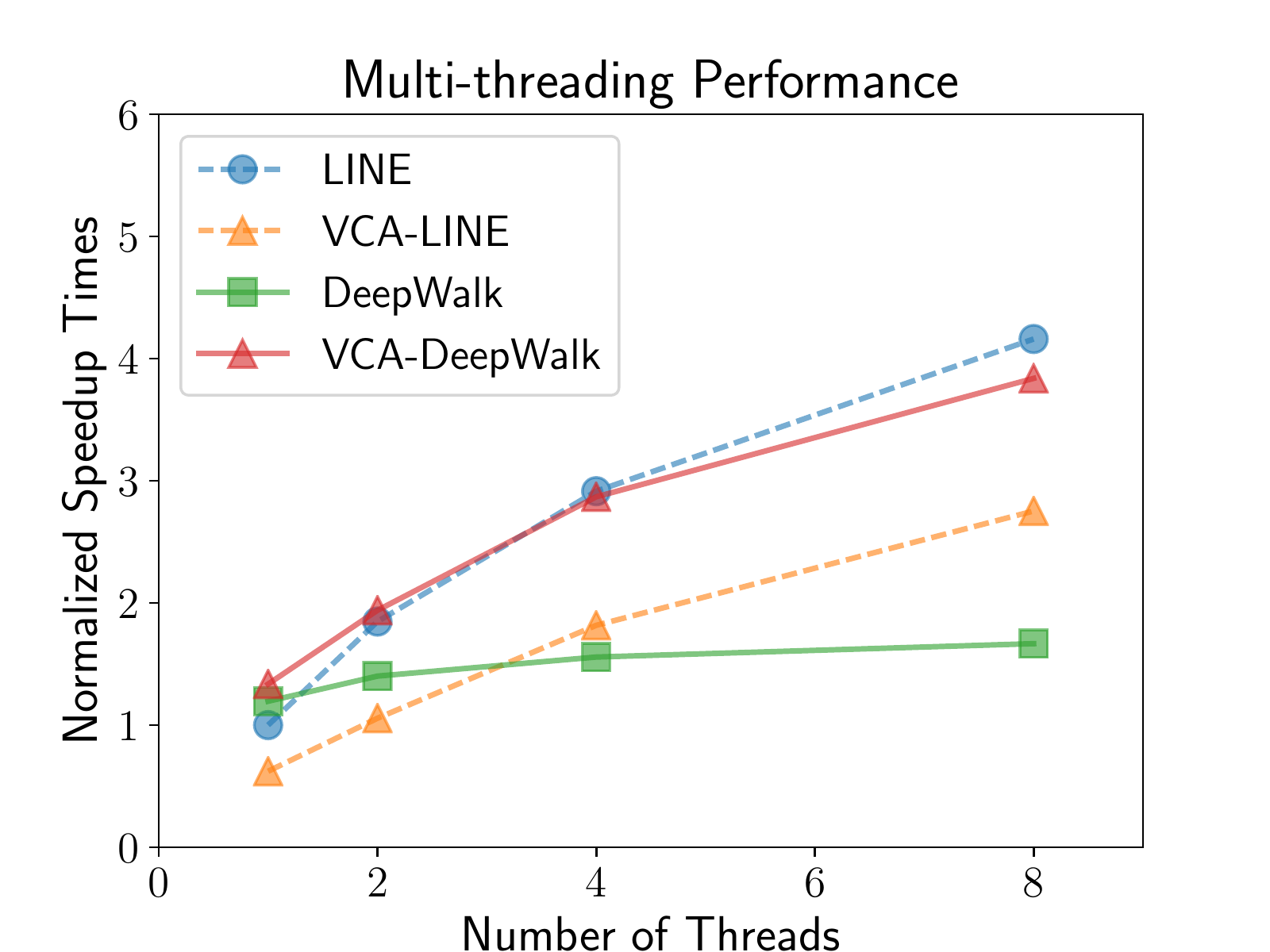}
\caption{Normalized speedup times of using multi-threads.}
\label{fig:time}
\end{figure}

Due to that the training pairs can be sampled in parallel and the representations can be updated simultaneously, we further report the relative speedup times among different implementations.  Consider that the original LINE algorithm is design for the large scale network, we normalize all the values using the division by the LINE algorithm with single thread, as shown in Figure~\ref{fig:time}.  Although the LINE algorithm is around two times faster than VCS-LINE, it has to be executed two times for obtaining the 1st and 2nd representations.  The bottleneck of original DeepWalk algorithm is the pre-computation of the walks, whereas our VCS-DeepWalk can sample a pair and update the representations at once. 


\section{Conclusion}
\label{sec_conclusion}

We have presented in this paper a novel weighted vertex-context sampling method for building information network embedding. In the experiments, the proposed VCS-based algorithms can serve as an example framework for building a weighted network embedding method. We show that integrating the edges weights with the conventional network embedding methods leads to a better performance. Moreover, the VCS-based approaches retain the modeling  flexibility with low time and space complexity. In the future work, we plan to deeply investigate how to redesign the weights and how to explore the local and global network structure by using the weighted vertex sampling methods. Moreover, we plan to extend this sampling method to the other kinds of models such as neural network-based and matrix factorization-based models.

\bibliographystyle{ACM-Reference-Format}
\bibliography{sigproc} 

\end{document}